\documentclass{article}
\usepackage{graphicx}
\usepackage{tabularx}
\usepackage[hyphens]{url}
\begin{document}

\title{\bf Malware Detection using Machine Learning and Deep Learning}
\author{\footnotesize Hemant Rathore, Swati Agarwal, Sanjay K. Sahay and Mohit Sewak \\
{\footnotesize\it BITS, Pilani, Dept. of CS \& IS, Goa Campus, Goa, India} \\
{\footnotesize\it Email: hemantr, swatia, ssahay, p20150023@goa.bits-pilani.ac.in}}

\date{}

\maketitle

\begin{abstract}
Research shows that over the last decade, malware have been growing exponentially, causing substantial financial losses to various organizations. Different anti-malware companies have been proposing solutions to defend attacks from these malware. The velocity, volume, and the complexity of malware are posing new challenges to the anti-malware community. Current state-of-the-art research shows that recently, researchers and anti-virus organizations started applying machine learning and deep learning methods for malware analysis and detection. We have used opcode frequency as a feature vector and applied unsupervised learning in addition to supervised learning for malware classification. The focus of this tutorial is to present our work on detecting malware with ($1$) various machine learning algorithms and ($2$) deep learning models. Our results show that the Random Forest outperforms Deep Neural Network with opcode frequency as a feature. Also in feature reduction, Deep Auto-Encoders are overkill for the dataset, and elementary function like Variance Threshold perform better than others.  In addition to the proposed methodologies, we will also discuss the additional issues and the unique challenges in the domain, open research problems, limitations, and future directions.
\vspace*{0.1cm}
~
\end{abstract}

%
%
%
\section{Introduction}
In the digital age, malware have impacted a large number of computing devices. The term malware come from \textbf{mal}icious soft\textbf{ware} which are designed to meet the harmful intent of a malicious attacker. Malware can compromise computers/smart devices, steal confidential information, penetrate networks, and cripple critical infrastructures, etc. These programs include viruses, worms, trojans, spyware, bots, rootkits, ransomware, etc. According to Computer Economics\footnote{\url{https://www.computereconomics.com/article.cfm?id=1225}}, financial loss due to malware attack has grown quadruple from \$3.3 billion in 1997 to \$13.3 billion in 2006. Every few years the definition of \textit{Year of Mega Breach} has to be recalibrated based on attacks performed in that particular year. Recently in 2016, \textit{WannaCry ransomware attack\footnote{\url{https://www.cbsnews.com/news/wannacry-ransomware-attacks-wannacry-virus-losses}}} crippled the computers of more than 150 countries, doing financial damage to different organizations.  In 2016, Cybersecurity Ventures\footnote{\url{https://cybersecurityventures.com/hackerpocalypse-cybercrime-report-2016}} estimated the total damage due to malware attacks was \$3 trillion in 2015 and is expected to reach \$6 trillion by 2021.

Antivirus software (such as Norton, McAfee, Avast, Kaspersky, AVG, Bitdefender, etc.) is a major line of defense for malware attacks. Traditionally, an antivirus software used the signature-based method for malware detection. Signature is a short sequence of bytes which can be used to identify known malware. But the signature-based detection system cannot provide security against zero-day attacks. Also, malware generation toolkits like Zeus \cite{zeus} can generate thousands of variant of the same malware by using different obfuscation techniques. Signature generation is often a human-driven process which is bound to become infeasible with the current malware growth.

In the past few years, researchers and anti-malware communities have reported using machine learning and deep learning based methods for designing malware analysis and detection system. We surveyed these systems and divided the existing literature into two lines of research. (1) \textbf{feature extraction and feature reduction}: In malware analysis, features can be generated in two different ways: static analysis and dynamic analysis. In static analysis, features are extracted without executing the code whereas in dynamic analysis features are derived while running the executable. Ye et al. \cite{ye2007imds} used Windows API calls obtained from the static analysis as they can reflect true intent or behavior of an attacker. Their experiments show that few API calls like \textit{OpenProcess, CloseHandle, CopyFileA} etc. always co-occur in malicious executables. Raff et al. \cite{zak2017can} concluded that byte level n-gram could gather a lot of information about maliciousness from the code section as compared to portable executable header or import sections in a binary file. Strings also contain crucial semantic details, and they can often reflect the attacker’s real intent and goals. Studies show that in a particular malware family, sample executables often share a similar group of opcodes \cite{ye2010automatic}. Also, few opcodes are more dominant in malicious files as compare to benign executables which can act as a distinguisher. During malware analysis often features vector become extensively large, and it can have a negative impact during modeling. Literature shows various feature selection methods like document frequency \cite{moskovitch2008unknown}, information gain \cite{masud2011cloud}, max-relevance algorithm \cite{ye2008intelligent} have been used in various malware detection systems.  Azar \cite{yousefi2017autoencoder} performed feature reduction using auto-encoders (in turn reducing the memory requirement) and applied various classification algorithms to achieve higher accuracy. David et al. \cite{david2015deepsign} used a deep stack of de-noising auto-encoders implemented as deep belief network to generate the reduced feature set. (2) \textbf{Building Classification Models}: After feature extraction each file can be represented as a feature vector which can be used by the classification algorithm to build a model for malware detection. Firdausi et al. \cite{firdausi2010analysis} used naive bayes, J48, decision tree, k-nearest neighbor, multi-level perceptron and support vector machine on features extracted (using dynamic analysis) and achieved the highest accuracy of 96.8\% with J48. Moskovitch \cite{moskovitch2008unknown} generated feature vectors with the byte n-gram method and applied feature selection based on document frequency and gain ratio. They reported highest accuracy by selecting top 300 5-gram terms with decision tree and artificial neural network. In 2013, Santos et al. \cite{santos2013} generated a combined feature vector from the static analysis (sequence of opcode frequency) and dynamic features (system call, exception, etc.) from a sample of 1000 malicious and 1000 benign files. Hardy et al. \cite{hardy2016dl4md} in 2016 used Windows API calls as features with stacked autoencoder for malware detection and achieved an accuracy of 96.85\%.

\section{Experimental Setup}
\begin{figure}
	\centering
	\includegraphics[scale=.59]{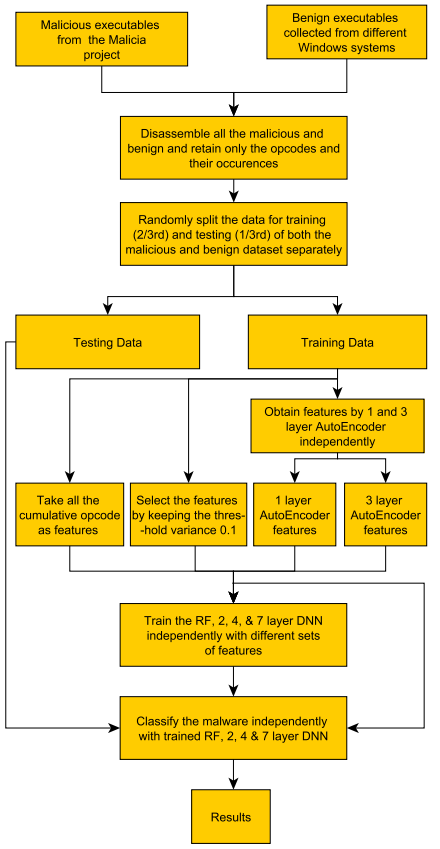}
	\caption{Flowchart for the classification of malware with different sets of features. (Source: Sewak et al. \cite{sewak2018comparison})}
	\label{flowchart}
\end{figure}

{\label{Experimental Setup}}
We formulate the problem of malware analysis and detection as a binary classification problem where malware and benign are the two classes. Figure \ref{flowchart} shows the proposed approach is a multi-step process consisting of various phases performing several tasks: collection of the dataset, disassembling of executable files, feature extraction, dimension reduction, building classification models, and empirical analysis of the results based on different metrics. We discuss each of these phases in the following subsections.

\subsection{Dataset}{\label{Dataset}}
To conduct our experiments, we gathered malware and benign executables from different sources. We downloaded malware samples from an open source repository known as Malicia Project\footnote{\url{http://malicia-project.com/}}. In Malicia Project, Nappa et al. \cite{nappa2013driving} have collected $11,688$ malware samples on Windows platform belonging to a total of $55$ different malware families The data collection is performed over a span of $11$ months ($07/03/2012$ to $25/03/2013$) from more than $500$ drive-by download servers also known as exploit servers. Typically these servers are deployed for a lifetime of $16$ hours while some servers even operated for months to spread the malware files. Many malware executables in the dataset will connect to the internet without user consent to perform some cybercrime operation. Most of the malicious executable will also repack themselves on an average of $5.4$ times in a day to evade the antivirus signature-based detection system. Thus opcode frequency as a feature can be an excellent measure to detect these malware.

To collect benign executable samples for our dataset, we gathered default files installed in different Windows operating system. VirusTotal\footnote{\url{https://www.virustotal.com/}} is an anti-virus aggregator that can be used to check whether an executable is malicious or benign. We declare a sample as non-malicious/benign when all the anti-virus from virustotal.com declares it as harmless. We combine the malware and benign executable files downloaded from different sources (Malicia and Windows) and use it as our experimental dataset. Thus the dataset contains $11,688$ malware and $2,819$ benign executable files.

\subsection{Disassembling of Malicious and Benign Executables}
As discussed in section \ref{Dataset} our data set consist of $14,507$ executable files. To generate the features, we disassemble them by converting an executable file (.exe) to assembly code (.asm). We used object dump utility which is a part of GNU Binutils package\footnote{\url{https://www.gnu.org/software/binutils/}}. During disassembling few executable files were found to be corrupted or encrypted thus those files were removed from the dataset. Finally, we used $2,819$ benign and $11,308$ malware executables to generate the feature vector and to build the classification model.

\subsection{Creation of Feature Vector space}{\label{Creation of Feature Vector space}}
In any machine learning algorithm, the feature vector is a critical component. We generate our feature vector by the static analysis of executable files. In static analysis, discriminatory attributes are collected without the execution of code. Literature shows that various static attribute such as Windows API calls \cite{ye2007imds} \cite{ye2017survey}, strings \cite{ye2017survey}, opcode \cite{sharma2016effective} \cite{sahay2016grouping}, and control flow graph \cite{ye2017survey}, etc. are used to separate the malicious and benign executables. We used opcode frequency as a discriminatory feature. Firstly an exhaustive feature list called as \textit{master opcode list} of $1,600$ unique opcodes was created. We future generate a  feature vector where rows represent the file name, and columns represent the opcode frequency. Each entry in the vector space represents the number of occurrence of a particular opcode in that file. Finally, the vector space of $2819 \times 1600$ for benign and $11308 \times 1600$ for malware executables was generated.

\subsection{Other issues}
Since there is a significant difference between the number of malware ($11,308$) and benign executables ($2,819$) in our dataset, thus it will lead to class imbalance problem. Various methods are available to solve class imbalance problem like random sampling (oversampling/undersampling), cluster-based sampling, ADASYN \cite{he2008adasyn}, etc. We used Adaptive Synthetic sampling approach for imbalanced learning (ADASYN) which is an oversampling method for minority class tuples. It synthetically generates data points of minority class based on the k-nearest neighbor algorithm.

As discussed in section \ref{Creation of Feature Vector space}, our dataset contains a large number of features and executable files thus we used cross-validation to generalize our model to an independent dataset. We used 3-fold cross validation in all our experiments. In rotation estimation (a.k.a. cross-validation) data is split into three equal parts where two blocks are used to training the model, and remaining one block is used for testing. The above exercise is done three times to accommodate all possible combinations.

\section{Modelling Malware Detection}{\label{Modelling Malware Detection}}
As discussed in section \ref{Experimental Setup}, malware detection is a binary classification problem. After disassembling the executable samples (malware/benign), successfully generating the feature vectors and using ADASYN, the next steps are dimensionality reduction and then finally building the classification models.

\subsection{Dimensionality Reduction}{\label{Dimensionality Reduction}}
In statistics and machine learning, dimensionality reduction is a process of reducing the number of features under consideration. Our feature vector suffers from the curse of dimensionality since the total number of the unique opcode is $1,600$. When we further analyzed our feature set, we found that for few opcodes the corresponding frequency is zero since the particular opcodes are deprecated. Also for few opcodes, the count was relatively less because they are platform specific and the platform is deprecated. A model created on a dataset suffering from the curse of dimensionality will take a longer time to train and is inefficient in space complexity as well. To choose an optimal number of features we are using different variants of dimensionality reduction methods.
 
\begin{figure}[!tbp]
	\centering
	\begin{minipage}[b]{0.45\textwidth}
		\includegraphics[width=1.0\linewidth]{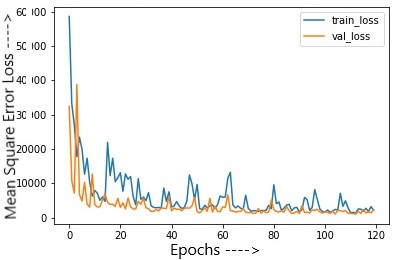}
		\caption{Plot for AE-1L shows mean squared error loss (y-axis) for training and validation across 120 epochs (x-axis) (Source: Sewak et al. \cite{sewak2018investigation})}
		\label{Sample_Plot_AE_Training}
	\end{minipage}
	\hfill
	\begin{minipage}[b]{0.45\textwidth}
		\includegraphics[width=1.0\linewidth]{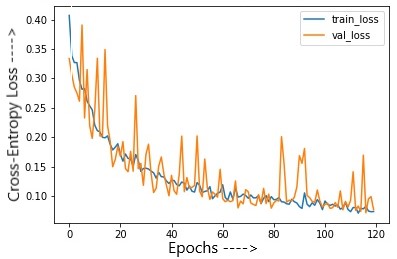}
		\caption{Plot for DNN-2L shows cross entropy loss (y-axis) for training and validation across 120 epochs (x-axis) (Source: Sewak et al. \cite{sewak2018investigation})}
		\label{Sample_Plot_DNN_Training}
	\end{minipage}
\end{figure}

\begin{enumerate}
	\item {\textit{None:}} In this method all the opcodes are taken into account for building a classification model without using any feature reduction. We use this as a baseline for different feature reduction methods.\\
	\item {\textit{Variance Threshold:}} It is a method used to remove the features with low variations. We have removed the attributes with a variance of less than 0.1 assuming they have less prediction power.\\
	\item {\textit{Auto-Encoders:}} In deep learning auto-encoders are unsupervised learning methods which require only feature vector (opcode frequency), and not class labels for dimensionality reduction.\\
	\begin{enumerate}
		\item A single layer auto-encoder (Non Deep Auto-Encoder), also referred to as AE-1L which contain one encoder layer and a decoder layer.\\
		\item A 3-layer stacked auto-encoder(Deep Auto-Encoder), also referred to as AE-3L which contain three encoders followed by three decoders.\\
	\end{enumerate}

	For our experiments, all the auto-encoders use Exponential Linear Unit (ELU) function at all the layers except in the last layer which uses linear activation function. In AE-1L, the input directly connects to bottleneck layer which in turn link to the output layer. In both the auto-encoder (AE-1L and AE-3L) models, the bottleneck layer consists of 32 ELU nodes. Thus the architecture of AE-1L is (Input-32-Output) where bottleneck layer will behave as both encoder and decoder. In case of AE-3L where encoder consists of two additional hidden layers connected in sequential order containing 128 and 64 nodes respectively. Similarly, AE-3L decoder comprised of two hidden layers of similar width but connected in reverse order. Thus architecture of AE-3L will be (Input-128-64-32-64-128-Output). For training of both the auto-encoders (AE-1L and AE-3L), the mean square error is used as a loss function over a batch size of 64 samples. Instead of using standard stochastic gradient we have used Adam optimizer \cite{kingma2014adam} to train a batch over 120 epochs. The figure (\ref{Sample_Plot_AE_Training}) shows the training and validation loss for AE-1L during a complete cycle. The plot shows mean squared error loss (y-axis) for training and validation which are converging around 120 epoch (x-axis).
\end{enumerate}

\subsection{Building the learning model}

In this paper, we used both machine learning and deep learning based approaches to build the classification models. Based on learning methods we divided our work into two case studies: (1) model based on the Random Forest (RF).  In the previous studies \cite{sharma2016effective} \cite{sahay2016grouping} conducted on the Malicia dataset \cite{nappa2013driving}, we found that tree-based classifier performs better as compared to other classifiers while among tree based classifier RF outperforms others. Thus we finally choose RF from the set of standard classifiers. (2) models based on deep learning.

\begin{enumerate}
	\item Deep Neural Network using two hidden layers (DNN-2L)
	\item Deep Neural Network using four hidden layers (DNN-4L)
	\item Deep Neural Network using seven hidden layers (DNN-7L)
\end{enumerate}

We designed multiple models of different depths to learn features at the different level of abstraction. In DNNs all the hidden layers contain ELU activation function except the last. Since malware detection is a binary classification problem, the last layer comprises of softmax activation (sigmoid) function. All the DNNs contain Adam optimizer \cite{kingma2014adam} instead of gradient decent since in general, they have faster convergence rate. Also, we used cross entropy loss function and to avoid overfitting problems we used a dropout rate of $0.1$. In DNN-2L, the two hidden layers contain 1024 and 32 nodes respectively. DNN-4L contain four layers with $2^{12-2i}$ nodes in each layer. Thus DNN-4L hidden layers contains (1024, 256,64,16) nodes. The DNN-7L has seven layers with $2^{11-i}$ nodes in $i^{th}$ hidden layer. Thus DNN-7L hidden layer contain (1024, 512, 256, 128, 64, 32, 16) nodes. Figure \ref{Sample_Plot_DNN_Training} shows the training and validation loss for DNN-2L for a complete cycle of 120 epochs. In this figure, both training and validation loss are gradually decreasing as the model parameters are getting trained in each epoch and finally converged around 120 epoch. Also, something training loss is more than validation loss which is counterintuitive but is it because of the drop-out rate (0.1) during the training cycle.

\section{Results}

In this section, we will discuss the experimental results obtained after feature reduction (refer  section \ref{Creation of Feature Vector space}) with classification models (refer section \ref{Modelling Malware Detection}) using various evaluation metrics (accuracy, recall, selectivity, and precision).

\begin{table}[]
	\centering
	\caption{Results with Features Reduction, Classification Models, Accuracy,  Recall /True Positive Rate (TPR),  Selectivity /True Negative Rate (TNR), Precision /Positive Predictive Value (PPV) (Source: Sewak et al. \cite{sewak2018comparison})}
	\begin{tabularx}{\textwidth}{|>{\centering\arraybackslash}X|>{\centering\arraybackslash}X|>{\centering\arraybackslash}X|>{\centering\arraybackslash}X|>{\centering\arraybackslash}X|>{\centering\arraybackslash}X|>{\centering\arraybackslash}X|}
		\hline
		\textbf{Features} & \textbf{Classifiers} & \textbf{Accuracy} & \textbf{TPR} & \textbf{TNR} & \textbf{PPV} \\
		\hline
		\hline
		None & RF & 99.74 & 99.48 & 100.0 & 100.0 \\
		VT & RF & 99.78 & 99.59 & 99.97 & 99.97  \\
		AE-1L & RF & 99.41 & 98.86 & 99.97 & 99.97  \\
		AE-3L & RF & 99.36 & 98.72 & 100.0 & 100.0  \\
		\hline
		None & DNN-2L & 97.79 & 96.33 & 99.26 & 99.24  \\
		VT & DNN-2L & 98.84 & 98.32 & 99.37 & 99.37  \\
		AE-1L & DNN-2L & 96.95 & 94.57 & 99.37 & 99.34  \\
		AE-3L & DNN-2L & 96.25 & 93.75 & 98.79 & 98.74  \\
		\hline
		None & DNN-4L & 97.42 & 95.38 & 99.48 & 99.46  \\
		VT & DNN-4L & 98.69 & 97.96 & 99.42 & 99.42  \\
		AE-1L & DNN-4L & 98.99 & 98.29 & 99.70 & 99.70  \\
		AE-3L & DNN-4L & 97.16 & 98.61 & 95.68 & 95.85  \\
		\hline
		None & DNN-7L & 96.15 & 99.05 & 93.20 & 93.66  \\
		VT & DNN-7L & 96.20 & 98.89 & 93.48 & 93.89  \\
		AE-1L & DNN-7L & 98.99 & 98.61 & 99.81 & 99.81  \\
		AE-3L & DNN-7L & 93.60 & 87.97 & 99.31 & 99.23  \\
		\hline
	\end{tabularx}
	\break
	\label{table2}
\end{table} 

Table \ref{table2} reveals that for different feature reduction methods we found that VT (combined with RF) based attribute reduction achieved the highest accuracy of 99.78\% which is marginally higher than no reduction (None and RF) 99.74\% in the feature set. AE-1L performed better than deeper Auto-Encoder (AE-3L) and obtained the highest accuracy (99.41\%) with RF. AE-3L based reduction performed lowest in all the methods. Highest True Positive Rate (TPR) of 99.59\% was archived by VT (and RF) followed by None, and highest True Negative Rate (TNR) of 100\% was achieved by no feature reduction (None and RF).

Table \ref{table2} shows that among different classification models, RF outperformed the deep learning models and achieved the highest accuracy of 99.7\% (RF and VT). RF again produced the second highest accuracy with no feature reduction. Between different deep learning models, DNN-3L and DNN-7L both combined with AE-1L attained an accuracy of 98.99\%. Highest TPR and TNR were produced by RF with VT and None as feature reduction respectively.

\section{Conclusion}

In the last few years malware have become a significant threat. Classical defense mechanism (like signature-based malware detection) used by anti-virus will fail to cope up new age malware challenges. In this paper, we have modeled malware analysis and detection as machine learning and deep learning problem. We have used best practices in building these models (like cross-validation, fixing class imbalance problem, etc.). We expertly handled the curse of dimensionality by using various feature reduction methods (None, AE-1L and AE-3L). Finally, we compared the models build using RF and DNN (DNN-2L, DNN-4L, and DNN-7L).  

Based on our results random forest outperforms all the three deep neural network models in malware detection. We achieved the highest accuracy of 99.78\% with random forest and variance threshold which is an improvement of 1.26\% on previously reported the best accuracy. Also in feature reduction, variance threshold outplayed auto-encoders in improving the model performance. Another significant contribution of our investigation is a comparison of different combinations of auto-encoder (of depth 1 and 3) and deep neural network  (of depth 2, 4 and 7) for malware detection. To our surprise, the best result did not come from any of the deep learning models which indicates that deep leaning may be overkill for Malicia dataset and the trained models are moving towards overfitting. 

The same models can be used to detect more complex malware (polymorphic and metamorphic) in the future. Further, it will be interesting to see the effectiveness of other deep learning techniques like recurrent neural network, long short-term memory, etc. for malware detection.

%
%
%
\section*{References}
\bibliography{bda2018}
\bibliographystyle{plain}

\end{document}